# Experimental Basis for Special Relativity in the Photon Sector

Daniel Y. Gezari

NASA/Goddard Space Flight Center, ExoPlanets and Stellar Astrophysics Laboratory,
Code 667, Greenbelt, MD  20771
*and*
American Museum of Natural History, Astrophysics Dept., New York, NY  10024

*daniel.y.gezari@nasa.gov*

ABSTRACT

A search of the literature reveals that none of the five new optical effects predicted by the special theory of relativity have ever been observed to occur in nature.  In particular, the speed of light ($c$) has never been measured directly with a moving detector to validate the invariance of $c$ to motion of the observer, a necessary condition for the Lorentz invariance of $c$.  The invariance of $c$ can now only be inferred from indirect experimental evidence. It is also not widely recognized that essentially all of the experimental support for special relativity in the photon sector consists of null results. The experimental basis for special relativity in the photon sector is summarized, and concerns about the completeness, integrity and interpretation of the present body of experimental evidence are discussed.

## 1.  INTRODUCTION

One of the most reassuring things we know about modern physics is that the special theory of relativity has faced a century of experimental challenges, and passed every test.  This is generally understood to mean that every aspect of special relativity has been tested and validated, beyond any doubt.  But all it really means is that every aspect of special relativity *that has been tested* has passed the test.  This prompts the question, what has been tested and what has not?

Contrary to the popular view, a search of the literature reveals that the experimental basis for the special theory of relativity in the photon sector is not robust.  Special relativity assumes or predicts eight new physical effects, three in the matter sector and five in the photon sector (Einstein 1905).  The three new matter effects are time dilation, mass increase and $E = mc^2$.  The five new photon effects are:

   invariance of the speed of light ($c$) to motion of the observer
   Lorentz-FitzGerald length contraction
   relativistic Doppler effect
   relativistic stellar aberration
   relativistic source brightening

Surprisingly, none of the five new optical effects assumed or predicted by special relativity have ever been observed to occur in nature or demonstrated in the laboratory. Principal among the unobserved effects is the invariance of $c$ to motion of the observer, the tacit assumption underlying all of the predictions of special relativity in the matter and photon sectors.  Because none of the new optical effects have ever been directly observed, the majority of tests of special relativity in the photon sector investigated secondary or implied effects. It is also not widely recognized that essentially all of the optical tests of special relativity produced null results (Section 3). That is, with the exception of four problematic experiments testing second-order Doppler effects (discussed in Section 3.5), there is no positive experimental evidence supporting special relativity in the photon sector.

The local Lorentz invariance of $c$ can now only be inferred from observations of moving sources, symmetry arguments, and the apparent null results of ether drift and speed-of-light isotropy experiments, but there are serious difficulties with this view: Observations of moving sources cannot discriminate between special relativity and the classical ether hypothesis (Section 2.1), the equivalence of source and observer motions has not been established experimentally (Section 2.1), the well-known ether drift experiments (*e.g.*, Michelson and Morley 1887) have recently come into question (Section 3.3), and resonant cavity isotropy experiments would be subject to those and additional concerns (Section 3.4).

## 2. PREDICTED SPECIAL RELATIVISTIC OPTICAL EFFECTS

The present experimental evidence for each of the five new optical effects predicted by the special theory of relativity can be summarized as follows:

### 2.1. Invariance of c

There are two necessary conditions for the local Lorentz invariance of $c$: invariance to motion of the source and invariance to motion of the observer. Satisfaction of these two conditions is both necessary and sufficient to validate the invariance of $c$. Invariance to motion of the emitting source – Einstein's second postulate has been convincingly validated experimentally (Section 3.1). But conspicuously absent from the experimental record is any published attempt to directly measure the speed of light with a moving detector to test the invariance of $c$ to motion of the observer.

The experimental validation of the invariance of $c$ is plagued by misconceptions and errors of interpretation. There is a common misconception that Einstein's second postulate says that $c$ is invariant to 'motion of the source and motion of the observer' and it is incorrectly presented this way in most textbooks. But the second postulate says nothing about the observer: "Light is propagated in empty space with a definite velocity $c$ which is independent of the state of motion of the emitting body" (Einstein 1905). The second postulate was not a new idea in 1905 and it is not unique to special relativity (recall that the classical wave theory of light also holds that $c$ is invariant to motion of the emitting source). So observations of moving sources cannot discriminate between special relativity and the old ether hypothesis, and do not favor one over the other. Of course, it could be argued that experiments with moving sources and moving observers should be equivalent and indistinguishable, so the second postulate would apply to the observer as well as to the source. But in other phenomena involving propagating light (*e.g.*, the Doppler effect in an optical medium, stellar aberration) motion of the source and motion of the observer have entirely independent consequences. To claim that source and observer motions are equivalent without experimental confirmation would be invoking the theory to validate itself. Observations of moving sources certainly cannot validate the universal Lorentz invariance of $c$ without observations with moving detectors, or at least experimental validation of the equivalence of source and observer motions for propagating light, and

### 2.2. Lorentz Length Contraction

Lorentz length contraction has never been observed to occur in nature, and there are no published reports of any formal experimental attempt to test it directly. Length contraction was originally thought to be a real physical deformation of the atomic lattice structure of solid matter. Later the model was relaxed to invoke only the appearance of contraction. Today, length contraction is considered to be a real optical effect, but one that is, in practice, unobservable because of a breakdown in simultaneity.

### 2.3. Relativistic Doppler Effect

No published experiment or observation has ever succeeded in resolving the relativistic departure of the longitudinal Doppler effect from its classical limit. The functional form of the relativistic Doppler effect is mathematically elegant but physically unrealistic: it is the geometric mean of the classical moving source and moving observer Doppler expressions. For all velocities $0 < v < 0.5c$ the relativistic Doppler shift is about half the classical moving source Doppler shift, and twice the classical moving observer shift. There would seem to be no physical mechanism to account for this peculiar behavior (Gezari 2010).

The relativistic Doppler effect implies the existence of two additional second-order optical effects: a slight asymmetry in the observed *longitudinal* Doppler effect for equal positive and negative velocities along the line-of-sight (investigated by Ives and Stilwell 1938, 1941, and Saathoff et al. 2003), and a *transverse* Doppler effect in light – a residual blueshift when relative motion is exactly perpendicular to the line-of-sight (investigated by Hasselkemp et al. 1979, and Kundig 1969). These four experiments testing second-order Doppler phenomena are the only positive results among all experimental tests of special relativity in the photon sector, however, all are problematic (as discussed in Section 3.5).

### 2.4. Relativistic Stellar Aberration

Relativistic stellar aberration has never been resolved from its classical counterpart because sufficiently high observer velocities have never been achieved. Furthermore, the aberration of starlight only results from motion of the observer – it is not produced by motion of the source; if it was the effect would be obvious in observations of nearby, fast-moving binary systems, such as Jupiter and its moon Io. For instance, if two observers were separated by a large distance in free space and at rest in the same inertial frame, and one of them went into uniform motion transverse to the line-of-sight between them, only one of them would observe the other's apparent position to be aberrated as a result of the net relative motion. But all observable effects in special relativity depend only on relative motion, so the asymmetrical nature of stellar aberration violates special relativity. However, the matter is by no means resolved; an interesting review and discussion was presented by Liebscher and Brosche (1998).

### 2.5. Relativistic Source Brightening

No published experiment or observation has ever reported the relativistic brightening of a light source measured by an approaching observer. Special relativity predicts that the intensity of a light source will become infinitely bright when measured by an observer approaching that source at $v \to c$. (This effect might be expected to have a classical counterpart, such as $I \to 2I$ as $v \to c$, however, there is no description of a classical brightening effect to be found in any physics textbook). Relativistic brightening is not the same as synchrotron forward beaming or relativistic stellar aberration. Occasionally astronomical phenomena such as asymmetric jets associated with supermassive black holes, AGN or blazars are attributed to relativistic brightening, but this is really only supposition.

## 3. EXPERIMENTAL TESTS IN THE PHOTON SECTOR

When Albert Einstein introduced the special theory of relativity in 1905 there had been only the most basic, static measurements of the speed of light (*e.g.*, Rømer 1676, Fizeau 1849, Foucault 1862, Michelson 1878, Newcomb 1886) and none of them considered the consequences of source or observer motion, or gave any indication that the speed of light might be invariant. The Michelson-Morley (1887) experiment had failed to detect the ether, which was perplexing and prompted theoretical attempts by FitzGerald (1889), Poincare (1900) and Lorentz (1904) to salvage the ether hypothesis. But Einstein did not refer to the Michelson-Morley experiment and apparently was not motivated by it (Brush 1999, Norton 2004). His purpose in proposing the special theory of relativity was to preserve Maxwellian electrodynamics. If $c$ was invariant, then Maxwell's equations would be covariant, and the question of the ether would be superfluous. Special relativity was pure speculation and wishful thinking in 1905. It remained for everything about the theory to be observed in nature or demonstrated in the laboratory. But in the first half of the 20$^{th}$ Century only seven experiments involving propagating light specifically testing special relativity were published, three between 1910 and 1914 that tested the second postulate by observations of moving sources (Section 3.1), and four between 1927 and 1933 that were improvements on the Michelson-Morley (1887) experiment (Section 3.3). During the same period only six primitive particle kinematics experiments were performed, all of which were published between 1906 and 1915 when particle physics was in its infancy. However, general relativity (Einstein 1916) successfully accounted for the precession of the perihelion of Mercury and predicted the gravitational bending of starlight, which increased confidence in Einstein's general approach. By the time modern instrumentation became available in the 1960's special relativity had already become part of the fabric of quantum mechanics and modern particle physics. A definitive historical study by Brush (1999) suggests that the early acceptance of special relativity was based more on its mathematical beauty than on the strength of the available experimental evidence, and the fact that it had been embraced by a few influential physicists (*e.g.*, Planck; Pauli 1921).

The most significant experiments testing special relativity published through 1990 have been discussed in a definitive review by Zhang (1999), and a convenient listing of all experiments published through 2007 has been compiled by Roberts (2008). Of perhaps 200 experiments testing special relativity reported in the professional literature at least sixty were tests in the photon sector.

### 3.1. Tests of the Second Postulate

The early acceptance of special relativity was based in large part on the validation of the second postulate by observations of visible binary stars (*e.g.*, Comstock 1910, De Sitter 1913, Zurhellen 1914). However, it was later recognized (Fox 1962) that all of the binary star observations in visible light were invalid because interstellar scintillation corrupts measurements over distances greater than ~1 parsec (*i.e.*, for all visible stars). The second postulate was eventually validated by the uniformity of the timing signature of the of the binary x-ray pulsar system Her X1 (Brecher 1977), by the measurement of the speed of gamma rays emitted by fast pions (Alvaeger et al. 1964, Filipas and Fox 1964) and by laboratory experiments with moving optical elements (Babcock and Bergmann 1964). All of the published observations of moving sources have produced null results, consistent with both special relativity and the classical ether hypothesis, but contradicting emission theories of light (*e.g.*, Ritz 1909).

### 3.2. Searches for a Light Medium

A group of experiments sought evidence of the ether by searching for variations in $c$ with frequency (*e.g.*, Nodland and Ralston 1997, Schaffer 1999). All produced null results. No variation of $c$ with frequency was observed, which argues against the existence of a physical light medium.

### 3.3. Ether Drift Tests

It is widely believed that all the early ether drift experiments (*e.g.*, Michelson and Morley 1887, Illingworth 1927, Kennedy and Thorndyke 1932 and Joos 1933) all produced null results, although Miller (1933) insisted that he consistently obtained ~8 *km/s* drift velocities over a period of 30 years, claims that were later discredited by Shankland, his former student (Shankland et al. 1955). Michelson and Morley (1887) also reported a net ~8 *km/s* drift velocity, which was widely interpreted as an upper limit and dismissed because it was much smaller than the expected ~30 *km/s* orbital velocity of the Earth. However, Consoli and Costanza (2003) recently pointed out that several of the well-known ether drift experiments actually produced small, positive drift velocities. They re-analyzed the original results of the experiments by Michelson-Morley (1887), Illingworth (1927), Miller (1933) and Joos (1933), correcting for Fresnel drag in the experiment medium (*i.e.*, in air, helium, rough lab vacuum) in the context of Lorentzian relativity, and obtained a corrected drift velocity of 204±36 *km/s* from the four experiments, which corresponds closely to the ~230 *km/s* orbital velocity of the Sun in the Galaxy. Comparable results and conclusions have also been presented by Munera (1998) and Cahill and Kitto (2002). These analyses suggest that the ether drift experiments may, in fact, show evidence of an absolute reference frame for the propagation of light.

### 3.4. Speed-of-Light Isotropy Tests

The classic resonant-cavity speed-of-light isotropy experiments (*e.g.*, Cederholm et al. 1959, Brillet and Hall 1979) have recently been repeated, and significantly improved upper limits have been set (*e.g.,* Kostelecky and Mewes 2002, Braxmaier et al. 2002, Muller et al. 2003, Muller 2005, Muller et al. 2006, Wolf et al. 2006). But isotropy experiments would be subject to the same concerns as those raised by Consoli and Costanza (2003) for the ether drift tests, and there are additional concerns for the integrity of the modern resonant cavity experiments. The resonant cavities are enclosed within metallic cryogenic/vacuum chambers so there is a very real possibility that the experiment apparatus itself could act to shield the sensitive elements from interaction with some absolute reference frame for light propagating in free space – should one exist – in the same sense that a Faraday cage shields an enclosed volume from external electromagnetic fields.

All isotropy experiments have been performed at the Earth's surface, below the atmosphere, and thus cannot be considered measurements made in "free space". Also, the physical characteristics of any putative preferred reference frame for the propagation of light are completely unknown, as are the possible interactions between such a frame and matter (*e.g.*, the Earth's atmosphere, the Solar wind, the zodiacal gas and dust cloud, etc.), or the extent of perturbation by zodiacal electric-magnetic-gravitational fields, so there is also fundamental concern that all terrestrial or solar-neighborhood experiments of this type may be compromised by particles and fields near the Earth's surface and in the solar neighborhood. These are all real concerns and they have not been formally addressed.

Finally, Zhang (1999) cautioned that all speed-of-light isotropy experiments have limited domains of applicability: "We have shown that the two-way velocity of light has been proven to be constant by means of these [isotropy] experiments and at the same time that the one-way velocity of light has been proved to be independent of the motion of the light source. It is needed to point out that 'independence' does not mean that either the one-way velocity of light is equal to the two-way velocity of light or the one-way velocity of light is isotropic."

### *3.5. Tests of Second-order Doppler Effects*

Ives and Stilwell (1938, 1941) performed a second-order Doppler experiment that was intended to be a demonstration of time dilation. They measured the longitudinal Doppler effect in red- and blue-shifted H$\beta$ line emission from hydrogen atoms moving at speeds up to $v = 0.01c$, simultaneously in the approaching and receding directions (essentially the same effect in neon was investigated by Kaivola et al. 1985, and McGowan et al. 1993). But these results and their interpretation are also problematic: While it is true that the observed red- and blue-shift are slightly asymmetrical in the relativistic case, they are also asymmetrical in the classical Doppler effect for moving sources, only this second-order effect is actually twice as large in the classical case than it is in the relativistic case, so detection would be easier in the classical case. The existence of an asymmetry does not necessarily mean that the observed effect was relativistic.

Kundig (1963) and Hasselkemp, Mondry and Scharmann (1979) claimed to have directly observed the relativistic transverse Doppler effect, however, the results and interpretation of both works are problematic. Kundig's Mossbauer rotor experiment was recently re-analyzed by Kholmetskii et al. (2008). Hasselkemp et al. did not initially detect a transverse Doppler effect, but claimed a detection after correcting for a possible misalignment. Furthermore, it is a little-known fact that a classical transverse Doppler effect is observed in acoustics, which would suggest an analog in the classical wave theory light, although no such effect is described in textbooks.

### *3.6. Other Relevant Experiments*

Georges Sagnac built a closed-path interferometer that detected its own rotation in the laboratory and announced that he had discovered the lumineferous ether (Sagnac 1913). Michelson and Gale (1925) built an evacuated 1000 x 2000 *ft* closed-path interferometer using the same principle and with it detected the rotation of the Earth. Sagnac effect fiberoptic gyros are used in all modern aerospace navigation systems.

It would seem that elements of the classical Sagnac effect conflict directly with special relativity, however, the prevailing view is that the rotating instrument is a non-inertial system to which special relativity does not apply (as first argued by Langevin 1921). The argument goes further to say that an observer viewing the rotating experiment from any inertial frame would be permitted under the rules of special relativity to measure relative speeds that differed from $c$, so the apparent speeds $c + v$ and $c - v$ of the counter-propagating beams in the instrument frame would still be consistent with special relativity. However, recently Wang et al. (2003, 2004) demonstrated the Sagnac effect in a non-rotating, inertial reference frame using a fiber optic linear motion sensor (FOLMS) interferometer. They showed that the light travel time in a straight optical fiber in inertial motion has a first-order dependence on the fiber speed in the local stationary frame, just as the light travel time in a rotating Sagnac effect fiber optic gyro has a first-order dependence on the tangential rotation speed of the fiber. The effect was obtained using both solid and hollow (air core) fibers. If the Sagnac effect can be produced by inertial motion then the rules of special relativity would have to be applied after all, and the linear Sagnac experiment would violate special relativity.

One of the most familiar arguments for the validity of special relativity is that the Global Positioning System (GPS) navigation system would not achieve high accuracy without making special relativistic corrections. The principal correction cited is a first-order timing adjustment to compensate for signal propagation time variations arising from the motions of the satellites and ground receiver in the local Earth-centered Earth-fixed (ECEF) frame due to the Sagnac effect. But the Sagnac effect is a purely classical, first-order effect that has somehow been incorrectly re-classified in this application as a special relativistic effect (see, *e.g.*, Allen, Weiss and Ashby 1985; Ashby 2002). So a first-order timing correction must be made for the accurate performance of the GPS system, just the amount attributable to the velocity component of the satellite constellation along the line-of-sight in the ECEF frame. The fact that this first-order timing correction is required at all is in direct conflict with special relativity.

Wolf and Petit (1997) tested the isotropy of the one-way speed of light by analyzing the GPS satellite network timing signal database and found no dependency on source-receiver motion, which they interpreted as evidence that $c$ was isotropic. However, Wolf and Petit noted that the data set they analyzed had been pre-processed and corrected for the Sagnac effect (in this case, a first-order change in the time of flight of radio signals between satellites and receivers), and justified making this correction by claiming that the Sagnac effect was a relativistic effect that was "second-order in $c$" and therefore that the correction had negligible consequences in their analysis. But the correction that was made was first-order in the velocity component of the line-of-sight between each satellite and the ground receiver in the ECEF frame. Thus, while the affect of the first-order line-of-sight velocity component of the GPS satellites is clearly evident in the raw timing signal data, this first-order component had been removed from the dataset that Wolf and Petit analyzed to show that $c$ was isotropic.

## 4. DISCUSSION

It is troubling that there are no unambiguous, positive experimental results in the photon sector to support the local Lorentz invariance of $c$. Null results are useful, but not as compelling as positive detections because they introduce various additional elements of uncertainty. A null result can be obtained if the experiment is not well-conceived or the instrumentation is not well-designed, if the detector is not sensitive to the intended effect, if the observation is not performed correctly, if the results are not analyzed or interpreted correctly, if the experimental test does not actually apply, or if natural phenomena corrupt the observation. Furthermore, many of the experimental null results and upper limits that are described as "consistent with special relativity" would be more correctly characterized as "not inconsistent with special relativity within the experimental uncertainties". Note that many of the same experiments that are consistent with special relativity are also consistent with classical physics (*e.g.*, all moving source observations testing the second postulate).

Ultimately, any concerns about the validity of a theory can only be resolved by experiment. It would be much more straightforward, and more convincing, to simply measure the speed of light directly with a moving detector that was controlled or actively monitored by the observer. Two simple experiments could test the invariance of $c$ to motion of the observer: 1) a direct measurement of the one-way speed of light with two detectors moving as a pair in the laboratory using femtosecond optical pulse timing, and 2) a measurement of the speed of light by timing of laser pulses propagating between the Earth and a retro-reflector on the surface of the Moon. In both experiments the detector module and timing electronics are at rest in the observer's frame, so any first-order change in the time-of-flight could only be due to a change in the speed of the pulse relative to the detector module.

## 5. CONCLUSION

Considering the weakness of the present experimental support for the invariance of $c$ – the fact that observations of moving sources cannot discriminate between special relativity and the old ether hypothesis, the absence of speed-of-light measurements with moving detectors, the lack of experimental validation of the equivalence of source and observer motions, doubts about the interpretation of the classical ether-drift experiments, concerns about the applicability of the modern isotropy experiments, and the fact that all of the unambiguous tests of special relativity in the photon sector have produced null results – it cannot yet be claimed that the local Lorentz invariance of $c$ has been convincingly validated by observation or experiment. It would be prudent to critically re-examine and strengthen the present experimental basis for the special theory of relativity in the photon sector. At least one of the five relativistic optical effects predicted by the special theory of relativity should be confirmed by direct observation; the most significant of these would be the invariance of $c$ to motion of the observer. To this end we have made a two-way lunar laser ranging measurement of the speed of light with a moving detector (Gezari 2009) and we are pursuing one-way laser ranging observations outside the Earth's atmosphere (Gezari et al. 2010) as well as ultra-fast pulse timing measurements in the laboratory (Gezari et al. 2010).

Acknowledgments

I am grateful to Bob Woodruff for many inspiring discussions and valuable suggestions that led to important insights. Special thanks to Bill Danchi, Bill Oegerle and Jennifer Wisemann for their objective support of this effort at Goddard. This research was supported by the National Aeronautics and Space Administration.